# A new algorithm for multiplying two Dirac numbers

Aleksandr Cariow[1], Galina Cariowa[1]

[1] Faculty of Computer Sciences and Information Technologies, Żołnierska 52
71-210 Szczecin, Poland
{atariov, gtariova}@wi.zut.edu.pl
tel. +48 91 4495573

**Abstract.** In this work a rationalized algorithm for Dirac numbers multiplication is presented. This algorithm has a low computational complexity feature and is well suited to FPGA implementation. The computation of two Dirac numbers product using the naïve method takes 256 real multiplications and 240 real additions, while the proposed algorithm can compute the same result in only 88 real multiplications and 256 real additions. During synthesis of the discussed algorithm we use the fact that Dirac numbers product may be represented as vector-matrix product. The matrix participating in the product has unique structural properties that allow performing its advantageous decomposition. Namely this decomposition leads to significant reducing of the computational complexity.

## 1. Introduction

Recently hypercomplex numbers [1] are used in various fields of data processing including digital signal and image processing, machine graphics, telecommunications and especially in public key cryptography [2-10]. The most popular are quaternions, octonions and sedenions [1]. Perhaps the less popular are the Pauli, Kaluza and Dirac numbers [11]. This numbers are mostly used in solving different physical problems in electrodynamics, field theory, etc. Anyway, hypercomplex arithmetic is a very important issue in modern data processing applications.

Among other operations in hypercomplex arithmetic, multiplication is the most time consuming one. The reason for this is, because the addition of $N$-dimensional hypercomplex numbers only requires $N$ real additions; the multiplication of these numbers already requires $N(N-1)$ real additions and $N^2$ real multiplication. It is easy to see that the increasing of dimensions of hypercomplex number increases the computational complexity of the multiplication. Therefore, reducing the computational complexity of the multiplication of hypercomplex numbers is an important scientific and engineering problem.

Several efficient algorithms for the multiplication of hypercomplex numbers have been reported in the literature [12-17]. Our previous work [12] proposed an algorithm for computing product of two Dirac numbers which has lower computational complexity compared with the direct (schoolbook) method of computations. In this paper we propose another algorithm for this purpose. Compared with our previous algorithm, the proposed algorithm has lower multiplicative complexity.

## 2. Preliminary Remarks

A Dirac number is defined as follows [12]:

$$d = d_0 + \sum_{n=1}^{15} d_n i_n \qquad (1)$$

where $d_0$ and $\{d_n\}, n=1,...,15$ - are real numbers, and $\{i_n\}, n=1,...,15$ - are the imaginary units that commute with real numbers during multiplication. At that $i_1$, $i_2$, $i_3$, $i_4$ - the main imaginary units, and the remaining imaginary units are composite ones and are expressed in terms of the main imaginary units by the formula:

$$i_s = i_p i_q \cdots i_r, \text{ where } 1 \le p < q \cdots < r \le n.$$

All products of imaginary units on each other are entirely determined by the predetermined rules for multiplication main imaginary units on each other:

$$i_p^2 = \varepsilon_p, \; i_q i_p = \alpha_{pq} i_p i_q, \; p < q, \; p,q = 1,2,...,n, \text{ where } \varepsilon_p, \; \alpha_{pq} \text{ are equal to } -1, 1, \text{ or } 0.$$

The results of all possible products of the Dirac numbers imaginary units can be summarized in the following table [12]:

| × | 1 | $i_1$ | $i_2$ | $i_3$ | $i_4$ | $i_5$ | $i_6$ | $i_7$ | $i_8$ | $i_9$ | $i_{10}$ | $i_{11}$ | $i_{12}$ | $i_{13}$ | $i_{14}$ | $i_{15}$ |
|---|---|---|---|---|---|---|---|---|---|---|---|---|---|---|---|---|
| 1 | 1 | $i_1$ | $i_2$ | $i_3$ | $i_4$ | $i_5$ | $i_6$ | $i_7$ | $i_8$ | $i_9$ | $i_{10}$ | $i_{11}$ | $i_{12}$ | $i_{13}$ | $i_{14}$ | $i_{15}$ |
| $i_1$ | $i_1$ | 1 | $i_5$ | $i_6$ | $i_7$ | $i_2$ | $i_3$ | $i_4$ | $i_{11}$ | $i_{12}$ | $i_{13}$ | $i_8$ | $i_9$ | $i_{10}$ | $i_{15}$ | $i_{14}$ |
| $i_2$ | $i_2$ | $-i_5$ | $-1$ | $i_8$ | $i_9$ | $i_1$ | $-i_{11}$ | $-i_{12}$ | $-i_3$ | $-i_4$ | $i_{14}$ | $i_6$ | $i_7$ | $-i_{15}$ | $-i_{10}$ | $i_{13}$ |
| $i_3$ | $i_3$ | $-i_6$ | $-i_8$ | $-1$ | $i_{10}$ | $i_{11}$ | $i_1$ | $-i_{13}$ | $i_2$ | $-i_{14}$ | $-i_4$ | $-i_5$ | $i_{15}$ | $i_7$ | $i_9$ | $-i_{12}$ |
| $i_4$ | $i_4$ | $-i_7$ | $-i_9$ | $-i_{10}$ | $-1$ | $i_{12}$ | $i_{13}$ | $i_1$ | $i_{14}$ | $i_2$ | $i_3$ | $-i_{15}$ | $-i_5$ | $-i_6$ | $-i_8$ | $i_{11}$ |
| $i_5$ | $i_5$ | $-i_2$ | $-i_1$ | $i_{11}$ | $i_{12}$ | 1 | $-i_8$ | $-i_9$ | $-i_6$ | $-i_7$ | $i_{15}$ | $i_3$ | $i_4$ | $-i_{14}$ | $-i_{13}$ | $i_{10}$ |
| $i_6$ | $i_6$ | $-i_3$ | $-i_{11}$ | $-i_1$ | $i_{13}$ | $i_8$ | 1 | $-i_{10}$ | $i_5$ | $-i_{15}$ | $-i_7$ | $-i_2$ | $i_{14}$ | $i_4$ | $i_{12}$ | $-i_9$ |
| $i_7$ | $i_7$ | $-i_4$ | $-i_{12}$ | $-i_{13}$ | $-i_1$ | $i_9$ | $i_{10}$ | 1 | $i_{15}$ | $i_5$ | $i_6$ | $-i_{14}$ | $-i_2$ | $-i_3$ | $-i_{11}$ | $i_8$ |
| $i_8$ | $i_8$ | $i_{11}$ | $i_3$ | $-i_2$ | $i_{14}$ | $i_6$ | $-i_5$ | $i_{15}$ | $-1$ | $i_{10}$ | $-i_9$ | $-i_1$ | $i_{13}$ | $-i_{12}$ | $-i_4$ | $-i_7$ |
| $i_9$ | $i_9$ | $i_{12}$ | $i_4$ | $-i_{14}$ | $-i_2$ | $i_7$ | $-i_{15}$ | $-i_5$ | $-i_{10}$ | $-1$ | $i_8$ | $-i_{13}$ | $-i_1$ | $i_{11}$ | $i_3$ | $i_6$ |
| $i_{10}$ | $i_{10}$ | $i_{13}$ | $i_{14}$ | $i_4$ | $-i_3$ | $i_{15}$ | $i_7$ | $-i_6$ | $i_9$ | $-i_8$ | $-1$ | $i_{12}$ | $-i_{11}$ | $-i_1$ | $-i_2$ | $-i_5$ |
| $i_{11}$ | $i_{11}$ | $i_8$ | $i_6$ | $-i_5$ | $i_{15}$ | $i_3$ | $-i_2$ | $i_{14}$ | $-i_1$ | $i_{13}$ | $-i_{12}$ | $-1$ | $i_{10}$ | $-i_9$ | $-i_7$ | $-i_4$ |
| $i_{12}$ | $i_{12}$ | $i_9$ | $i_7$ | $-i_{15}$ | $-i_5$ | $i_4$ | $-i_{14}$ | $-i_2$ | $-i_{13}$ | $-i_1$ | $i_{11}$ | $-i_{10}$ | $-1$ | $i_8$ | $i_6$ | $i_3$ |
| $i_{13}$ | $i_{13}$ | $i_{10}$ | $i_{15}$ | $i_7$ | $-i_6$ | $i_{14}$ | $i_4$ | $-i_3$ | $i_{12}$ | $-i_{11}$ | $-i_1$ | $i_9$ | $-i_8$ | $-1$ | $-i_5$ | $-i_2$ |
| $i_{14}$ | $i_{14}$ | $-i_{15}$ | $-i_{10}$ | $i_9$ | $-i_8$ | $i_{13}$ | $-i_{12}$ | $i_{11}$ | $-i_4$ | $i_3$ | $-i_2$ | $i_7$ | $-i_6$ | $i_5$ | 1 | $-i_1$ |
| $i_{15}$ | $i_{15}$ | $-i_{14}$ | $-i_{13}$ | $i_{12}$ | $-i_{11}$ | $i_{10}$ | $-i_9$ | $i_8$ | $-i_7$ | $i_6$ | $-i_5$ | $i_4$ | $-i_3$ | $i_2$ | $i_1$ | $-1$ |

Consider the problem of multiplying two Dirac numbers:

$$d = d^{(1)} \cdot d^{(2)} = d_0 + \sum_{n=1}^{15} i_n d_n, \; d^{(1)} = a_0 + \sum_{n=1}^{15} i_n a_n, \; d^{(2)} = b_0 + \sum_{n=1}^{15} i_n b_n.$$

The operation of multiplication of Dirac numbers can be represented more compactly in the form of vector-matrix product:

$$\mathbf{Y}_{16 \times 1} = \mathbf{B}_{16} \mathbf{X}_{16 \times 1} \tag{2}$$

where

$$\mathbf{X}_{16 \times 1} = [a_0, a_1, ..., a_{15}]^T, \; \mathbf{Y}_{16 \times 1} = [d_0, d_1, ..., d_{15}]^T,$$

$$\mathbf{B}_{16} = \begin{bmatrix}
b_0 & b_1 & -b_2 & -b_3 & -b_4 & b_5 & b_6 & b_7 & -b_8 & -b_9 & -b_{10} & -b_{11} & -b_{12} & -b_{13} & b_{14} & -b_{15} \\
b_1 & b_0 & b_5 & -b_2 & b_6 & -b_3 & b_7 & -b_4 & -b_{11} & -b_{12} & -b_{13} & -b_8 & -b_9 & -b_{10} & -b_{15} & b_{14} \\
b_2 & b_5 & b_0 & -b_1 & b_8 & -b_{11} & b_9 & -b_{12} & -b_3 & -b_4 & -b_{14} & -b_6 & -b_7 & -b_{15} & -b_{10} & b_{13} \\
b_3 & b_6 & -b_8 & b_{11} & b_0 & -b_1 & b_{10} & -b_{13} & b_2 & b_{14} & -b_4 & b_5 & b_{15} & -b_7 & b_9 & -b_{12} \\
b_4 & b_7 & -b_9 & b_{12} & -b_{10} & b_{13} & b_0 & -b_1 & -b_{14} & b_2 & b_3 & -b_{15} & b_5 & b_6 & -b_8 & b_{11} \\
b_5 & b_2 & -b_1 & b_0 & -b_{11} & b_8 & -b_{12} & b_9 & -b_6 & -b_7 & -b_{15} & -b_3 & -b_4 & -b_{14} & b_{13} & -b_{10} \\
b_6 & b_3 & b_{11} & -b_8 & -b_1 & b_0 & -b_{13} & b_{10} & b_5 & b_{15} & -b_7 & b_2 & b_{14} & -b_4 & -b_{12} & b_9 \\
b_7 & b_4 & b_{12} & -b_9 & b_{13} & -b_{10} & -b_1 & b_0 & -b_{15} & b_5 & b_6 & -b_{14} & b_2 & b_3 & b_{11} & -b_8 \\
b_8 & b_{11} & b_3 & -b_6 & -b_2 & b_5 & -b_{14} & b_{15} & b_0 & b_{10} & -b_9 & b_1 & b_{13} & -b_{12} & -b_4 & b_7 \\
b_9 & b_{12} & b_4 & -b_7 & b_{14} & -b_{15} & -b_2 & b_5 & -b_{10} & b_0 & b_8 & -b_{13} & b_1 & b_{11} & b_3 & -b_6 \\
b_{10} & b_{13} & -b_{14} & b_{15} & b_4 & -b_7 & -b_3 & b_6 & b_9 & -b_8 & b_0 & b_{12} & -b_{11} & b_1 & -b_2 & b_5 \\
b_{11} & b_8 & -b_6 & b_3 & b_5 & -b_2 & b_{15} & -b_{14} & b_1 & b_{13} & -b_{12} & b_0 & b_{10} & -b_9 & b_7 & -b_4 \\
b_{12} & b_9 & -b_7 & b_4 & -b_{15} & b_{14} & b_5 & -b_2 & -b_{13} & b_1 & b_{11} & -b_{10} & b_0 & b_8 & -b_6 & b_3 \\
b_{13} & b_{10} & b_{15} & -b_{14} & -b_7 & b_4 & b_6 & -b_3 & b_{12} & -b_{11} & b_1 & b_9 & -b_8 & b_0 & b_5 & -b_2 \\
b_{14} & b_{15} & b_{10} & -b_{13} & -b_9 & b_{12} & b_8 & -b_{11} & b_4 & -b_3 & b_2 & b_7 & -b_6 & b_5 & b_0 & -b_1 \\
b_{15} & b_{14} & -b_{13} & b_{10} & b_{12} & -b_9 & -b_{11} & b_8 & b_7 & -b_6 & b_5 & b_4 & -b_3 & b_2 & -b_1 & b_0
\end{bmatrix},$$

The direct multiplication of two Dirac numbers requires 256 real multiplications and 240 real additions. Our previous paper [12] reported an algorithm for multiplication of two Dirac numbers with 128 real multiplications and 160 real additions. In this paper we introduce the new algorithm, which reduce multiplicative complexity to 88 real multiplications at the cost of 96 extra real additions compared with our previous algorithm.

## 3. Synthesis of a rationalized algorithm for computing Dirac numbers product

First and foremost, we rearrange the columns of the matrix in the following order {1, 2, 3, 4, 6, 7, 9, 12, 11, 14, 15, 5, 16, 8, 10, and 13}. Next, we rearrange the rows of obtained matrix in the same manner. Next, let us multiply by (−1) every element in the lower half of the rows in the obtained matrix. As a result, we obtain the following matrix:

$$\tilde{\mathbf{B}}_{16} = \left[\begin{array}{cccccccc|cccccccc} b_0 & b_1 & -b_2 & -b_3 & b_5 & b_6 & -b_8 & -b_{11} & -b_{10} & -b_{13} & b_{14} & -b_4 & -b_{15} & b_7 & -b_9 & -b_{12} \\ b_1 & b_0 & b_5 & b_6 & -b_2 & -b_3 & -b_{11} & -b_8 & -b_{13} & -b_{10} & -b_{15} & b_7 & b_{14} & -b_4 & -b_{12} & -b_9 \\ b_2 & b_5 & b_0 & b_8 & -b_1 & -b_{11} & -b_3 & -b_6 & -b_{14} & -b_{15} & -b_{10} & b_9 & b_{13} & -b_{12} & -b_4 & -b_7 \\ b_3 & b_6 & -b_8 & b_0 & b_{11} & -b_1 & b_2 & b_5 & -b_4 & -b_7 & b_9 & b_{10} & -b_{12} & -b_{13} & b_{14} & b_{15} \\ b_5 & b_2 & -b_1 & -b_{11} & b_0 & b_8 & -b_6 & -b_3 & -b_{15} & -b_{14} & b_{13} & -b_{12} & -b_{10} & b_9 & -b_7 & -b_4 \\ b_6 & b_3 & b_{11} & -b_1 & -b_8 & b_0 & b_5 & b_2 & -b_7 & -b_4 & -b_{12} & -b_{13} & b_9 & b_{10} & b_{15} & b_{14} \\ b_8 & b_{11} & b_3 & -b_2 & -b_6 & b_5 & b_0 & b_1 & -b_9 & -b_{12} & -b_4 & -b_{14} & b_7 & b_{15} & b_{10} & b_{13} \\ b_{11} & b_8 & -b_6 & b_5 & b_3 & -b_2 & b_1 & b_0 & -b_{12} & -b_9 & b_7 & b_{15} & -b_4 & -b_{14} & b_{13} & b_{10} \\ \hline -b_{10} & -b_{13} & b_{14} & -b_4 & -b_{15} & b_7 & -b_9 & -b_{12} & -b_0 & -b_1 & b_2 & b_3 & -b_5 & -b_6 & b_8 & b_{11} \\ -b_{13} & -b_{10} & -b_{15} & b_7 & b_{14} & -b_4 & -b_{12} & -b_9 & -b_1 & -b_0 & -b_5 & -b_6 & b_2 & b_3 & b_{11} & b_8 \\ -b_{14} & -b_{15} & -b_{10} & b_9 & b_{13} & -b_{12} & -b_4 & -b_7 & -b_2 & -b_5 & -b_0 & -b_8 & b_1 & b_{11} & b_3 & b_6 \\ -b_4 & -b_7 & b_9 & b_{10} & -b_{12} & -b_{13} & b_{14} & b_{15} & -b_3 & -b_6 & b_8 & -b_0 & -b_{11} & b_1 & -b_2 & -b_5 \\ -b_{15} & -b_{14} & b_{13} & -b_{12} & -b_{10} & b_9 & -b_7 & -b_4 & -b_5 & -b_2 & b_1 & b_{11} & -b_0 & -b_8 & b_6 & b_3 \\ -b_7 & -b_4 & -b_{12} & -b_{13} & b_9 & b_{10} & b_{15} & b_{14} & -b_6 & -b_3 & -b_{11} & b_1 & b_8 & -b_0 & -b_5 & -b_2 \\ -b_9 & -b_{12} & -b_4 & -b_{14} & b_7 & b_{15} & b_{10} & b_{13} & -b_8 & -b_{11} & -b_3 & b_2 & b_6 & -b_5 & -b_1 & -b_0 \\ -b_{12} & -b_9 & b_7 & b_{15} & -b_4 & -b_{14} & b_{13} & b_{10} & -b_{11} & -b_8 & b_6 & -b_5 & -b_3 & b_2 & -b_1 & -b_0 \end{array}\right]$$

Then we can rewrite expression (2) in following form:

$$\mathbf{Y}_{16\times 1} = \mathbf{E}_{16}\mathbf{P}_{16}^{(2)}\tilde{\mathbf{B}}_{16}\mathbf{P}_{16}^{(1)}\mathbf{X}_{16\times 1} \qquad (3)$$

where

$$\mathbf{E}_{16} = diag(1,1,1,1,1,1,1,1,-1,-1,-1,-1,-1,-1,-1,-1),$$

$$\mathbf{P}_{16}^{(1)} = \begin{bmatrix} 1 & & & & & & & & & & & & & & & \\ & 1 & & & & & & & & & & & & & & \\ & & 1 & & & & & & & & & & & & & \\ & & & 1 & & & & & & & & & & & & \\ & & & & & 1 & & & & & & & & & & \\ & & & & & & 1 & & & & & & & & & \\ & & & & & & & & 1 & & & & & & & \\ & & & & & & & & & & & 1 & & & & \\ & & & & & & & & & & 1 & & & & & \\ & & & & & & & & & & & & 1 & & & \\ & & & & & & & & & & & & & 1 & & \\ & & & & 1 & & & & & & & & & & & \\ & & & & & & & & & & & & & & & 1 \\ & & & & & & & 1 & & & & & & & & \\ & & & & & & & & & 1 & & & & & & \\ & & & & & & & & & & & & 1 & & & \end{bmatrix},$$

$$\mathbf{P}_{16}^{(2)} = \begin{bmatrix} 1 & & & & & & & & & & & & & & & \\ & 1 & & & & & & & & & & & & & & \\ & & 1 & & & & & & & & & & & & & \\ & & & 1 & & & & & & & & & & & & \\ & & & & 1 & & & & & & & & & & & \\ & & & & & 1 & & & & & & & & & & \\ & & & & & & 1 & & & & & & & & & \\ & & & & & & & 1 & & & & & & & & \\ & & & & & & & & -1 & & & & & & & \\ & & & & & & & & & -1 & & & & & & \\ & & & & & & & & & & -1 & & & & & \\ & & & & & & & & & & & & & & & -1 \\ & & & & & & & & & & & -1 & & & & \\ & & & & & & & & & & & & -1 & & & \\ & & & & & & & & & & & & & -1 & & \\ & & & & & & & & & & & & & & -1 & \end{bmatrix}.$$

It is easy to see that $\tilde{\mathbf{B}}_{16}$ has the following structure:

$$\tilde{\mathbf{B}}_{16} = \begin{bmatrix} \mathbf{A}_8 & \mathbf{B}_8 \\ \mathbf{B}_8 & -\mathbf{A}_8 \end{bmatrix},$$

where

$$\mathbf{A}_8 = \begin{bmatrix} b_0 & b_1 & -b_2 & -b_3 & b_5 & b_6 & -b_8 & -b_{11} \\ b_1 & b_0 & b_5 & b_6 & -b_2 & -b_3 & -b_{11} & -b_8 \\ b_2 & b_5 & b_0 & b_8 & -b_1 & -b_{11} & -b_3 & -b_6 \\ b_3 & b_6 & -b_8 & b_0 & b_{11} & -b_1 & b_2 & b_5 \\ b_5 & b_2 & -b_1 & -b_{11} & b_0 & b_8 & -b_6 & -b_3 \\ b_6 & b_3 & b_{11} & -b_1 & -b_8 & b_0 & b_5 & b_2 \\ b_8 & b_{11} & b_3 & -b_2 & -b_6 & b_5 & b_0 & b_1 \\ b_{11} & b_8 & -b_6 & b_5 & b_3 & -b_2 & b_1 & b_0 \end{bmatrix},$$

$$\mathbf{B}_8 = \begin{bmatrix} -b_{10} & -b_{13} & b_{14} & -b_4 & -b_{15} & b_7 & -b_9 & -b_{12} \\ -b_{13} & -b_{10} & -b_{15} & b_7 & b_{14} & -b_4 & -b_{12} & -b_9 \\ -b_{14} & -b_{15} & -b_{10} & b_9 & b_{13} & -b_{12} & -b_4 & -b_7 \\ -b_4 & -b_7 & b_9 & b_{10} & -b_{12} & -b_{13} & b_{14} & b_{15} \\ -b_{15} & -b_{14} & b_{13} & -b_{12} & -b_{10} & b_9 & -b_7 & -b_4 \\ -b_7 & -b_4 & -b_{12} & -b_{13} & b_9 & b_{10} & b_{15} & b_{14} \\ -b_9 & -b_{12} & -b_4 & -b_{14} & b_7 & b_{15} & b_{10} & b_{13} \\ -b_{12} & -b_9 & b_7 & b_{15} & -b_4 & -b_{14} & b_{13} & b_{10} \end{bmatrix}.$$

As was shown in [18, 19], the matrix having such a structure can be efficiently factorized. This factorization reduces the number of multiplications by 25%. Than a computational procedure for the multiplication of Dirac numbers can be represented as follows:

$$\mathbf{Y}_{16\times 1} = \mathbf{E}_{16}\mathbf{P}_{16}^{(2)}(\mathbf{T}_{2\times 3} \otimes \mathbf{I}_8) diag\begin{bmatrix} \mathbf{A}_8 - \mathbf{B}_8 \\ -(\mathbf{A}_8 + \mathbf{B}_8) \\ \mathbf{B}_8 \end{bmatrix}(\mathbf{T}_{3\times 2} \otimes \mathbf{I}_8)\mathbf{P}_{16}^{(1)}\mathbf{X}_{16\times 1} \quad (4)$$

where $\mathbf{I}_N$ - is an identity $N \times N$ matrix, sign „$\otimes$"– denotes tensor product of two matrices [20],

$$\mathbf{T}_{2\times 3} = \begin{bmatrix} 1 & 0 & 1 \\ 0 & 1 & 1 \end{bmatrix}, \quad \mathbf{T}_{3\times 2} = \begin{bmatrix} 1 & 0 \\ 0 & 1 \\ 1 & 1 \end{bmatrix}$$

$$\mathbf{A}_8 - \mathbf{B}_8 = \left[\begin{array}{cccc|cccc}
b_0+b_{10} & b_1+b_{13} & -b_2-b_{14} & -b_3+b_4 & b_5+b_{15} & b_6-b_7 & -b_8+b_9 & -b_{11}+b_{12} \\
b_1+b_{13} & b_0+b_{10} & b_5+b_{15} & b_6-b_7 & -b_2-b_{14} & -b_3+b_4 & -b_{11}+b_{12} & -b_8+b_9 \\
b_2+b_{14} & b_5+b_{15} & b_0+b_{10} & b_8-b_9 & -b_1-b_{13} & -b_{11}+b_{12} & -b_3+b_4 & -b_6+b_7 \\
b_3+b_4 & b_6+b_7 & -b_8-b_9 & b_0-b_{10} & b_{11}+b_{12} & -b_1+b_{13} & b_2-b_{14} & b_5-b_{15} \\ \hline
b_5+b_{15} & b_2+b_{14} & -b_1-b_{13} & -b_{11}+b_{12} & b_0+b_{10} & b_8-b_9 & -b_6+b_7 & -b_3+b_4 \\
b_6+b_7 & b_3+b_4 & b_{11}+b_{12} & -b_1+b_{13} & -b_8-b_9 & b_0-b_{10} & b_5-b_{15} & b_2-b_{14} \\
b_8+b_9 & b_{11}+b_{12} & b_3+b_4 & -b_2+b_{14} & -b_6-b_7 & b_5-b_{15} & b_0-b_{10} & b_1-b_{13} \\
b_{11}+b_{12} & b_8+b_9 & -b_6-b_7 & b_5-b_{15} & b_3+b_4 & -b_2+b_{14} & b_1-b_{13} & b_0-b_{10}
\end{array}\right],$$

$$-(\mathbf{A}_8+\mathbf{B}_8) = \left[\begin{array}{cccc|cccc}
-b_0+b_{10} & -b_1+b_{13} & b_2-b_{14} & b_3+b_4 & -b_5+b_{15} & -b_6-b_7 & b_8+b_9 & b_{11}+b_{12} \\
-b_1+b_{13} & -b_0+b_{10} & -b_5+b_{15} & -b_6-b_7 & b_2-b_{14} & b_3+b_4 & b_{11}+b_{12} & b_8+b_9 \\
-b_2+b_{14} & -b_5+b_{15} & -b_0+b_{10} & -b_8-b_9 & b_1-b_{13} & b_{11}+b_{12} & b_3+b_4 & b_6+b_7 \\
-b_3+b_4 & -b_6+b_7 & b_8-b_9 & -b_0-b_{10} & -b_{11}+b_{12} & b_1+b_{13} & -b_2-b_{14} & -b_5-b_{15} \\ \hline
-b_5+b_{15} & -b_2+b_{14} & b_1-b_{13} & b_{11}+b_{12} & -b_0+b_{10} & -b_8-b_9 & b_6+b_7 & b_3+b_4 \\
-b_6+b_7 & -b_3+b_4 & -b_{11}+b_{12} & b_1+b_{13} & b_8-b_9 & -b_0-b_{10} & -b_5-b_{15} & -b_2-b_{14} \\
-b_8+b_9 & -b_{11}+b_{12} & -b_3+b_4 & b_2+b_{14} & b_6-b_7 & -b_5-b_{15} & -b_0-b_{10} & -b_1-b_{13} \\
-b_{11}+b_{12} & -b_8+b_9 & b_6-b_7 & -b_5-b_{15} & -b_3+b_4 & b_2+b_{14} & -b_1-b_{13} & -b_0-b_{10}
\end{array}\right].$$

Let we rearrange the columns of the matrix $\mathbf{A}_8 - \mathbf{B}_8$ in the following order: {1, 2, 4, 7, 5, 3, 8, 6}. After such permutation we multiply by (-1) every element of the sixth and eighth columns of the resulting matrix. The rows of the obtained matrix we rearrange in following order: {1, 7, 3, 4, 5, 6, 2, 8}. Then we multiply by (-1) every element of the sixth and seventh rows of the resulting matrix too. As a result, we obtain the following matrix:

$$\mathbf{B}'_8 = \left[\begin{array}{cccc|cccc}
b_0+b_{10} & b_1+b_{13} & -b_3+b_4 & -b_8+b_9 & b_5+b_{15} & b_2+b_{14} & -b_{11}+b_{12} & -b_6+b_7 \\
b_8+b_9 & b_{11}+b_{12} & -b_2+b_{14} & b_0-b_{10} & -b_6-b_7 & -b_3-b_4 & b_1-b_{13} & -b_5+b_{15} \\
b_2+b_{14} & b_5+b_{15} & b_8-b_9 & -b_3+b_4 & -b_1-b_{13} & -b_0-b_{10} & -b_6+b_7 & b_{11}-b_{12} \\
b_3+b_4 & b_6+b_7 & b_0-b_{10} & b_2-b_{14} & b_{11}+b_{12} & b_8+b_9 & b_5-b_{15} & b_1-b_{13} \\ \hline
b_5+b_{15} & b_2+b_{14} & -b_{11}+b_{12} & -b_6+b_7 & b_0+b_{10} & b_1+b_{13} & -b_3+b_4 & -b_8+b_9 \\
-b_6-b_7 & -b_3-b_4 & b_1-b_{13} & -b_5+b_{15} & b_8+b_9 & b_{11}+b_{12} & -b_2+b_{14} & b_0-b_{10} \\
-b_1-b_{13} & -b_0-b_{10} & -b_6+b_7 & b_{11}-b_{12} & b_2+b_{14} & b_5+b_{15} & b_8-b_9 & -b_3+b_4 \\
b_{11}+b_{12} & b_8+b_9 & b_5-b_{15} & b_1-b_{13} & b_3+b_4 & b_6+b_7 & -b_{10}+b_0 & b_2-b_{14}
\end{array}\right].$$

It is easy to see that the matrix $\mathbf{B}'_8$ possesses a structure that provides "good" factorization too.

$$\mathbf{B}'_8 = \left[\begin{array}{c|c} \mathbf{A}_4 & \mathbf{B}_4 \\ \hline \mathbf{B}_4 & \mathbf{A}_4 \end{array}\right],$$

where

$$\mathbf{A}_4 = \begin{bmatrix}
b_0+b_{10} & b_1+b_{13} & -b_3+b_4 & -b_8+b_9 \\
b_8+b_9 & b_{11}+b_{12} & -b_2+b_{14} & b_0-b_{10} \\
b_2+b_{14} & b_5+b_{15} & b_8-b_9 & -b_3+b_4 \\
b_3+b_4 & b_6+b_7 & b_0-b_{10} & b_2-b_{14}
\end{bmatrix}, \quad \mathbf{B}_4 = \begin{bmatrix}
b_5+b_{15} & b_2+b_{14} & -b_{11}+b_{12} & -b_6+b_7 \\
-b_6-b_7 & -b_3-b_4 & b_1-b_{13} & -b_5+b_{15} \\
-b_1-b_{13} & -b_0-b_{10} & -b_6+b_7 & b_{11}-b_{12} \\
b_{11}+b_{12} & b_8+b_9 & b_5-b_{15} & b_1-b_{13}
\end{bmatrix},$$

As follows from [18, 19], such block-structural form reduces the number of multiplications by 50%. This brings to the following factorization:

$$\mathbf{B'}_8 = (\mathbf{H}_2 \otimes \mathbf{I}_4) \frac{1}{2} (\mathbf{S}_4^{(0)} \oplus \mathbf{S}_4^{(1)}) (\mathbf{H}_2 \otimes \mathbf{I}_4) \qquad (5)$$

where sign " $\oplus$ " – denotes direct sum of two matrices and $\mathbf{H}_2 = \begin{bmatrix} 1 & 1 \\ 1 & -1 \end{bmatrix}$ is the $2 \times 2$ Hadamard matrix [20].

$$\mathbf{S}_4^{(0)} = \mathbf{A}_4 + \mathbf{B}_4, \quad \mathbf{S}_4^{(0)} = \left[ \begin{array}{cccc|cccc} b_0 + b_5 + b_{10} + b_{15} & b_1 + b_2 + b_{13} + b_{14} & -b_3 + b_4 - b_{11} + b_{12} & -b_6 + b_7 - b_8 + b_9 \\ -b_6 - b_7 + b_8 + b_9 & -b_3 - b_4 + b_{11} + b_{12} & b_1 - b_2 - b_{13} + b_{14} & b_0 - b_5 - b_{10} + b_{15} \\ \hline -b_1 + b_2 - b_{13} + b_{14} & -b_0 + b_5 - b_{10} + b_{15} & -b_6 + b_7 + b_8 - b_9 & -b_3 + b_4 + b_{11} - b_{12} \\ b_3 + b_4 + b_{11} + b_{12} & b_6 + b_7 + b_8 + b_9 & b_0 + b_5 - b_{10} - b_{15} & b_1 + b_2 - b_{13} - b_{14} \end{array} \right],$$

$$\mathbf{S}_4^{(1)} = \mathbf{A}_4 - \mathbf{B}_4, \quad \mathbf{S}_4^{(1)} = \left[ \begin{array}{cccc|cccc} b_0 - b_5 + b_{10} - b_{15} & b_1 - b_2 + b_{13} - b_{14} & -b_3 + b_4 + b_{11} - b_{12} & b_6 - b_7 - b_8 + b_9 \\ b_6 + b_7 + b_8 + b_9 & b_3 + b_4 + b_{11} + b_{12} & -b_1 - b_2 + b_{13} + b_{14} & b_0 + b_5 - b_{10} - b_{15} \\ \hline b_1 + b_2 + b_{13} + b_{14} & b_0 + b_5 + b_{10} + b_{15} & b_6 - b_7 + b_8 - b_9 & -b_3 + b_4 - b_{11} + b_{12} \\ b_3 + b_4 - b_{11} - b_{12} & b_6 + b_7 - b_8 - b_9 & b_0 - b_5 - b_{10} + b_{15} & -b_1 + b_2 + b_{13} - b_{14} \end{array} \right].$$

Next we rearrange the columns of the matrix $-(\mathbf{A}_8 + \mathbf{B}_8)$ in the following order: {1, 2, 4, 7, 5, 3, 8, 6}. Then we multiply by (-1) every element of the sixth and eighth columns of the resulting matrix. We rearrange the rows of the obtained matrix in the following order: {1, 7, 3, 4, 5, 6, 2, 8}. Then we multiply by (-1) every element of the sixth and eighth rows of the resulting matrix. As a result, we obtain the following matrix:

$$\mathbf{B}''_8 = \left[ \begin{array}{cccc|cccc} -b_0 + b_{10} & -b_1 + b_{13} & b_3 + b_4 & b_8 + b_9 & -b_5 + b_{15} & -b_2 + b_{14} & b_{11} + b_{12} & b_6 + b_7 \\ -b_1 + b_{13} & -b_0 + b_{10} & -b_6 - b_7 & b_{11} + b_{12} & b_2 - b_{14} & b_5 - b_{15} & b_8 + b_9 & -b_3 - b_4 \\ -b_3 + b_4 & -b_6 + b_7 & -b_0 - b_{10} & -b_2 - b_{14} & -b_{11} + b_{12} & -b_8 + b_9 & -b_5 - b_{15} & -b_1 - b_{13} \\ -b_8 + b_9 & -b_{11} + b_{12} & b_2 + b_{14} & -b_0 - b_{10} & b_6 - b_7 & b_3 - b_4 & -b_1 - b_{13} & b_5 + b_{15} \\ \hline -b_5 + b_{15} & -b_2 + b_{14} & b_{11} + b_{12} & b_6 + b_7 & -b_0 + b_{10} & -b_1 + b_{13} & b_3 + b_4 & b_8 + b_9 \\ b_2 - b_{14} & b_5 - b_{15} & b_8 + b_9 & -b_3 - b_4 & -b_1 + b_{13} & -b_0 + b_{10} & -b_6 - b_7 & b_{11} + b_{12} \\ -b_{11} + b_{12} & -b_8 + b_9 & -b_5 - b_{15} & -b_1 - b_{13} & -b_3 + b_4 & -b_6 + b_7 & -b_0 - b_{10} & -b_2 - b_{14} \\ b_6 - b_7 & b_3 - b_4 & -b_1 - b_{13} & b_5 + b_{15} & -b_8 + b_9 & -b_{11} + b_{12} & b_2 + b_{14} & -b_0 - b_{10} \end{array} \right]$$

It is easy to see that the matrix $\mathbf{B}''_8$ has the same type of structure as in the previous case:

$$\mathbf{B}''_8 = \begin{bmatrix} \mathbf{C}_4 & \mathbf{D}_4 \\ \mathbf{D}_4 & \mathbf{C}_4 \end{bmatrix},$$

where

$$\mathbf{C}_4 = \begin{bmatrix} -b_0 + b_{10} & -b_1 + b_{13} & b_3 + b_4 & b_8 + b_9 \\ -b_1 + b_{13} & -b_0 + b_{10} & -b_6 - b_7 & b_{11} + b_{12} \\ -b_3 + b_4 & -b_6 + b_7 & -b_0 - b_{10} & -b_2 - b_{14} \\ -b_8 + b_9 & -b_{11} + b_{12} & b_2 + b_{14} & -b_0 - b_{10} \end{bmatrix}, \quad \mathbf{D}_4 = \begin{bmatrix} -b_5 + b_{15} & -b_2 + b_{14} & b_{11} + b_{12} & b_6 + b_7 \\ b_2 - b_{14} & b_5 - b_{15} & b_8 + b_9 & -b_3 - b_4 \\ -b_{11} + b_{12} & -b_8 + b_9 & -b_5 - b_{15} & -b_1 - b_{13} \\ b_6 - b_7 & b_3 - b_4 & -b_1 - b_{13} & b_5 + b_{15} \end{bmatrix}.$$

Then we can write:

$$\mathbf{B}''_8 = (\mathbf{H}_2 \otimes \mathbf{I}_4) \frac{1}{2} (\mathbf{S}_4^{(2)} \oplus \mathbf{S}_4^{(3)}) (\mathbf{H}_2 \otimes \mathbf{I}_4) \qquad (6)$$

where

$$\mathbf{S}_4^{(2)} = \mathbf{C}_4 + \mathbf{D}_4, \quad \mathbf{S}_4^{(2)} = \left[ \begin{array}{cc|cc} -b_0 - b_5 + b_{10} + b_{15} & -b_1 - b_2 + b_{13} + b_{14} & b_3 + b_4 + b_{11} + b_{12} & b_6 + b_7 + b_8 + b_9 \\ -b_1 + b_2 + b_{13} - b_{14} & -b_0 + b_5 + b_{10} - b_{15} & -b_6 - b_7 + b_8 + b_9 & -b_3 - b_4 + b_{11} + b_{12} \\ \hline -b_3 + b_4 - b_{11} + b_{12} & -b_6 + b_7 - b_8 + b_9 & -b_0 - b_5 - b_{10} - b_{15} & -b_1 - b_2 - b_{13} - b_{14} \\ b_6 - b_7 - b_8 + b_9 & b_3 - b_4 - b_{11} + b_{12} & -b_1 + b_2 - b_{13} + b_{14} & -b_0 + b_5 - b_{10} + b_{15} \end{array} \right],$$

$$\mathbf{S}_4^{(3)} = \mathbf{C}_4 - \mathbf{D}_4, \quad \mathbf{S}_4^{(3)} = \begin{bmatrix} -b_0+b_5+b_{10}-b_{15} & -b_1+b_2+b_{13}-b_{14} & b_3+b_4-b_{11}-b_{12} & -b_6-b_7+b_8+b_9 \\ -b_1-b_2+b_{13}+b_{14} & -b_0-b_5+b_{10}+b_{15} & -b_6-b_7-b_8-b_9 & b_3+b_4+b_{11}+b_{12} \\ \hline -b_3+b_4+b_{11}-b_{12} & -b_6+b_7+b_8-b_9 & -b_0+b_5-b_{10}+b_{15} & b_1-b_2+b_{13}-b_{14} \\ -b_6+b_7-b_8+b_9 & -b_3+b_4-b_{11}+b_{12} & b_1+b_2+b_{13}+b_{14} & -b_0-b_5-b_{10}-b_{15} \end{bmatrix},$$

Lastly we rearrange the columns of the matrix $\mathbf{B}_8$ in the following order: {4, 2, 3, 8, 1, 6, 7, 5}. We rearrange the rows of the obtained matrix in the following order: {1, 6, 7, 5, 4, 2, 3, 8}. Then we multiply by (-1) every element of the fifth, sixth, seventh and eighth rows of the resulting matrix too. As a result, we obtain the following matrix:

$$\mathbf{B}_8''' = \begin{bmatrix} -b_4 & -b_{13} & b_{14} & -b_{12} & -b_{10} & b_7 & -b_9 & -b_{15} \\ -b_{13} & -b_4 & -b_{12} & b_{14} & -b_7 & b_{10} & b_{15} & b_9 \\ -b_{14} & -b_{12} & -b_4 & b_{13} & -b_9 & b_{15} & b_{10} & b_7 \\ -b_{12} & -b_{14} & b_{13} & -b_4 & -b_{15} & b_9 & -b_7 & -b_{10} \\ \hline -b_{10} & b_7 & -b_9 & -b_{15} & b_4 & b_{13} & -b_{14} & b_{12} \\ -b_7 & b_{10} & b_{15} & b_9 & b_{13} & b_4 & b_{12} & -b_{14} \\ -b_9 & b_{15} & b_{10} & b_7 & b_{14} & b_{12} & b_4 & -b_{13} \\ -b_{15} & b_9 & -b_7 & -b_{10} & b_{12} & b_{14} & -b_{13} & b_4 \end{bmatrix},$$

The matrix thus obtained has the following structure:

$$\mathbf{B}_8''' = \begin{bmatrix} \mathbf{E}_4 & \mathbf{F}_4 \\ \hline \mathbf{F}_4 & -\mathbf{E}_4 \end{bmatrix}, \quad \mathbf{E}_4 = \begin{bmatrix} -b_4 & -b_{13} & b_{14} & -b_{12} \\ -b_{13} & -b_4 & -b_{12} & b_{14} \\ -b_{14} & -b_{12} & -b_4 & b_{13} \\ -b_{12} & -b_{14} & b_{13} & -b_4 \end{bmatrix}, \quad \mathbf{F}_4 = \begin{bmatrix} -b_{10} & b_7 & -b_9 & -b_{15} \\ -b_7 & b_{10} & b_{15} & b_9 \\ -b_9 & b_{15} & b_{10} & b_7 \\ -b_{15} & b_9 & -b_7 & -b_{10} \end{bmatrix}.$$

Then we can write:

$$\mathbf{B}_8''' = (\mathbf{T}_{2\times 3} \otimes \mathbf{I}_4)(\mathbf{S}_4^{(4)} \oplus \mathbf{S}_4^{(5)} \oplus \mathbf{S}_4^{(6)})(\mathbf{T}_{3\times 2} \otimes \mathbf{I}_4) \qquad (7)$$

$$\mathbf{S}_4^{(4)} = \mathbf{E}_4 - \mathbf{F}_4, \quad \mathbf{S}_4^{(5)} = -(\mathbf{E}_4 + \mathbf{F}_4), \quad \mathbf{S}_4^{(6)} = \mathbf{F}_4.$$

$$\mathbf{S}_4^{(4)} = \begin{bmatrix} -b_4+b_{10} & -b_7-b_{13} & b_9+b_{14} & -b_{12}+b_{15} \\ b_7-b_{13} & -b_4-b_{10} & -b_{12}-b_{15} & -b_9+b_{14} \\ \hline b_9-b_{14} & -b_{12}-b_{15} & -b_4-b_{10} & -b_7+b_{13} \\ -b_{12}+b_{15} & -b_9-b_{14} & b_7+b_{13} & -b_4+b_{10} \end{bmatrix}, \quad \mathbf{S}_4^{(5)} = \begin{bmatrix} b_4+b_{10} & -b_7+b_{13} & b_9-b_{14} & b_{12}+b_{15} \\ b_7+b_{13} & b_4-b_{10} & b_{12}-b_{15} & -b_9-b_{14} \\ \hline b_9+b_{14} & b_{12}-b_{15} & b_4-b_{10} & -b_7-b_{13} \\ b_{12}+b_{15} & -b_9+b_{14} & b_7-b_{13} & b_4+b_{10} \end{bmatrix},$$

$$\mathbf{S}_4^{(6)} = \begin{bmatrix} -b_{10} & b_7 & -b_9 & -b_{15} \\ -b_7 & b_{10} & b_{15} & b_9 \\ \hline -b_9 & b_{15} & b_{10} & b_7 \\ -b_{15} & b_9 & -b_7 & -b_{10} \end{bmatrix}.$$

Combining (5), (6), and (7) and taking into account all manipulations with rows and columns in each matrix, we obtain a following vector-matrix procedure:

$$\mathbf{Y}_{16\times 1} = \mathbf{E}_{16}^{(1)} \mathbf{P}_{16}^{(2)} \mathbf{W}_{16\times 24}^{(1)} \mathbf{P}_{24}^{(4)} \mathbf{W}_{24\times 28}^{(2)} \mathbf{D}_{28}^{(2)} \tilde{\mathbf{W}}_{28\times 24}^{(2)} \mathbf{P}_{24}^{(3)} \tilde{\mathbf{W}}_{24\times 16}^{(1)} \mathbf{P}_{16}^{(1)} \mathbf{X}_{16\times 1} \qquad (8)$$

where

$$\mathbf{W}_{16\times 24}^{(1)} = \mathbf{T}_{2\times 3} \otimes \mathbf{I}_8, \quad \tilde{\mathbf{W}}_{24\times 16}^{(1)} = \mathbf{T}_{3\times 2} \otimes \mathbf{I}_8, \quad \mathbf{W}_{24\times 28}^{(2)} = (\mathbf{I}_2 \otimes (\mathbf{H}_2 \otimes \mathbf{I}_4)) \oplus (\mathbf{T}_{2\times 3} \otimes \mathbf{I}_4),$$

$$\widetilde{\mathbf{W}}_{28\times 24}^{(2)} = (\mathbf{I}_2 \otimes (\mathbf{H}_2 \otimes \mathbf{I}_4)) \oplus (\mathbf{T}_{3\times 2} \otimes \mathbf{I}_4), \quad \mathbf{D}_{28}^{(2)} = diag(\frac{1}{2}\mathbf{S}_4^{(0)}, \frac{1}{2}\mathbf{S}_4^{(1)}, \frac{1}{2}\mathbf{S}_4^{(2)}, \frac{1}{2}\mathbf{S}_4^{(3)}, \mathbf{S}_4^{(4)}, \mathbf{S}_4^{(5)}, \mathbf{S}_4^{(6)}).$$

$$\mathbf{P}_{24}^{(3)} = \begin{bmatrix}
1 & & & & & & & & & & & & & & & & & & & & & & & \\
& 1 & & & & & & & & & & & & & & & & & & & & & & \\
& & 1 & & & & & & & & & & & & & & & & & & & & & \\
& & & 1 & & & & & & & & & & & & & & & & & & & & \\
& & & & 1 & & & & & & & & & & & & & & & & & & & \\
& & & -1 & & & & & & & & & & & & & & & & & & & & \\
& & & & & & 1 & & & & & & & & & & & & & & & & & \\
& & & & & -1 & & & & & & & & & & & & & & & & & & \\
& & & & & & & 1 & & & & & & & & & & & & & & & & \\
& & & & & & & & 1 & & & & & & & & & & & & & & & \\
& & & & & & & & & 1 & & & & & & & & & & & & & & \\
& & & & & & & & & & 1 & & & & & & & & & & & & & \\
& & & & & & & & & & & 1 & & & & & & & & & & & & \\
& & & & & & & & & & -1 & & & & & & & & & & & & & \\
& & & & & & & & & & & & & 1 & & & & & & & & & & \\
& & & & & & & & & & & & -1 & & & & & & & & & & & \\
& & & & & & & & & & & & & & 1 & & & & & & & & & \\
& & & & & & & & & & & & & & & 1 & & & & & & & & \\
& & & & & & & & & & & & & & & & 1 & & & & & & & \\
& & & & & & & & & & & & & & & & & & & & & 1 & & \\
& & & & & & & & & & & & & & & & & 1 & & & & & & \\
& & & & & & & & & & & & & & & & & & & & & & 1 & \\
& & & & & & & & & & & & & & & & & & & & & & & 1 \\
& & & & & & & & & & & & & & & & & & & & 1 & & &
\end{bmatrix}$$

$$\mathbf{P}_{24}^{(4)} = \begin{bmatrix}
1 & & & & & & & & & & & & & & & & & & & & & & & \\
& & & & 1 & & & & & & & & & & & & & & & & & & & \\
& 1 & & & & & & & & & & & & & & & & & & & & & & \\
& & 1 & & & & & & & & & & & & & & & & & & & & & \\
& & & 1 & & & & & & & & & & & & & & & & & & & & \\
& & & & -1 & & & & & & & & & & & & & & & & & & & \\
-1 & & & & & & & & & & & & & & & & & & & & & & & \\
& & & & & & 1 & & & & & & & & & & & & & & & & & \\
& & & & & & & 1 & & & & & & & & & & & & & & & & \\
& & & & & & & & 1 & & & & & & & & & & & & & & & \\
& & & & & & & & & 1 & & & & & & & & & & & & & & \\
& & & & & & & & & & 1 & & & & & & & & & & & & & \\
& & & & & & & & & & & 1 & & & & & & & & & & & & \\
& & & & & & & & & & & & -1 & & & & & & & & & & & \\
& & & & & & & & & & & & & 1 & & & & & & & & & & \\
& & & & & & & & & & & & & & -1 & & & & & & & & & \\
& & & & & & & & & & & & & & & 1 & & & & & & & & \\
& & & & & & & & & & & & & & & & & & & 1 & & & & \\
& & & & & & & & & & & & & & & & & & & & 1 & & & \\
& & & & & & & & & & & & & & & & & 1 & & & & & & \\
& & & & & & & & & & & & & & & & & & -1 & & & & & \\
& & & & & & & & & & & & & & & & & & & & & -1 & & \\
& & & & & & & & & & & & & & & & & & & & & & -1 & \\
& & & & & & & & & & & & & & & & & & & & & & & -1
\end{bmatrix}$$

Unfortunately matrices $\mathbf{S}_4^{(0)}, \mathbf{S}_4^{(1)}, \mathbf{S}_4^{(2)}$ and $\mathbf{S}_4^{(3)}$ can no longer be effectively factorized. Their block composition is not conducive to a reduction of computational complexity. As for the matrix $\mathbf{S}_4^{(4)}$, its block structure after some modifications can be reduced to a convenient form. If we rearrange the columns and rows of the matrix $\mathbf{S}_4^{(4)}$ in the following order: {1, 2, 4, 3} and then multiply by (-1) every element of the last column and every element of the last row, we obtain the matrix

$$\widetilde{\mathbf{S}}_4^{(4)} = \begin{bmatrix} -b_4+b_{10} & -b_7-b_{13} & -b_{912}+b_{15} & -b_9-b_{14} \\ b_7-b_{13} & -b_4-b_{10} & -b_9+b_{14} & b_{12}+b_{15} \\ -b_{12}+b_{15} & -b_9-b_{14} & -b_4+b_{10} & -b_7-b_{13} \\ -b_9+b_{14} & b_{12}+b_{15} & b_7-b_{13} & -b_4-b_{10} \end{bmatrix} = \begin{bmatrix} \mathbf{A}_2 & \mathbf{B}_2 \\ \mathbf{B}_2 & \mathbf{A}_2 \end{bmatrix},$$

$$\mathbf{A}_2 = \begin{bmatrix} -b_4+b_{10} & -b_7-b_{13} \\ b_7-b_{13} & -b_4-b_{10} \end{bmatrix}, \quad \mathbf{B}_2 = \begin{bmatrix} -b_{12}+b_{15} & -b_9-b_{14} \\ -b_9+b_{14} & b_{12}+b_{15} \end{bmatrix},$$

whose structure has a "good" factorization property. Then we can write:

$$\widetilde{\mathbf{S}}_4^{(4)} = (\mathbf{H}_2 \otimes \mathbf{I}_2)\frac{1}{2}(\mathbf{S}_2^{(0)} \oplus \mathbf{S}_2^{(1)})(\mathbf{H}_2 \otimes \mathbf{I}_2) \qquad (9)$$

$$\mathbf{S}_2^{(0)} = \mathbf{A}_2 + \mathbf{B}_2, \quad \mathbf{S}_2^{(1)} = \mathbf{A}_2 - \mathbf{B}_2.$$

In turn, if we rearrange the columns and rows of the matrix $\mathbf{S}_4^{(5)}$ as {1, 2, 4, 3} and then multiply by (-1) every element of the last column and the every element of the last row, we obtain the matrix

$$\widetilde{\mathbf{S}}_4^{(5)} = \begin{bmatrix} b_4+b_{10} & -b_7+b_{13} & b_{12}+b_{15} & -b_9+b_{14} \\ b_7+b_{13} & b_4-b_{10} & -b_9-b_{14} & -b_{12}+b_{15} \\ b_{12}+b_{15} & -b_9+b_{14} & b_4+b_{10} & -b_7+b_{13} \\ -b_9-b_{14} & -b_{12}+b_{15} & b_7+b_{13} & b_4-b_{10} \end{bmatrix} = \begin{bmatrix} \mathbf{C}_2 & \mathbf{D}_2 \\ \mathbf{D}_2 & \mathbf{C}_2 \end{bmatrix},$$

$$\mathbf{C}_2 = \begin{bmatrix} b_4+b_{10} & -b_7+b_{13} \\ b_7+b_{13} & b_4-b_{10} \end{bmatrix}, \quad \mathbf{D}_2 = \begin{bmatrix} b_{12}+b_{15} & -b_9+b_{14} \\ -b_9-b_{14} & -b_{12}+b_{15} \end{bmatrix},$$

which has a "good" block structure too. We can write:

$$\widetilde{\mathbf{S}}_4^{(5)} = (\mathbf{H}_2 \otimes \mathbf{I}_2)\frac{1}{2}(\mathbf{S}_2^{(2)} \oplus \mathbf{S}_2^{(3)})(\mathbf{H}_2 \otimes \mathbf{I}_2) \qquad (10)$$

$$\mathbf{S}_2^{(2)} = \mathbf{C}_2 + \mathbf{D}_2, \quad \mathbf{S}_2^{(3)} = \mathbf{C}_2 - \mathbf{D}_2.$$

Let we rearrange columns of the matrix $\mathbf{S}_4^{(6)}$ in following way {1, 4, 3, 2}. Then multiply by (-1) every element of the last row, we obtain the matrix

$$\widetilde{\mathbf{S}}_4^{(6)} = \begin{bmatrix} -b_{10} & -b_{15} & -b_9 & b_7 \\ -b_7 & b_9 & b_{15} & b_{10} \\ -b_9 & b_7 & b_{10} & b_{15} \\ b_{15} & b_{10} & b_7 & -b_9 \end{bmatrix} = \begin{bmatrix} \mathbf{E}_2 & \mathbf{F}_2 \\ \mathbf{F}_2 & -\mathbf{E}_2 \end{bmatrix}, \quad \mathbf{E}_2 = \begin{bmatrix} -b_{10} & -b_{15} \\ -b_7 & b_9 \end{bmatrix}, \quad \mathbf{F}_2 = \begin{bmatrix} -b_9 & b_7 \\ b_{15} & b_{10} \end{bmatrix}.$$

Then we can write [18, 19]:

$$\widetilde{\mathbf{S}}_4^{(6)} = (\mathbf{T}_{2\times 3} \otimes \mathbf{I}_2)(\mathbf{S}_2^{(4)} \oplus \mathbf{S}_2^{(5)} \oplus \mathbf{F}_2)(\mathbf{T}_{3\times 2} \otimes \mathbf{I}_4) \qquad (11)$$

$$\mathbf{S}_2^{(4)} = \mathbf{E}_2 - \mathbf{F}_2, \quad \mathbf{S}_2^{(5)} = -(\mathbf{E}_2 + \mathbf{F}_2).$$

Combining (9), (10), and (11) and taking into account all manipulations with rows and columns in each matrix at this stage of synthesis of algorithm, we obtain a new updated matrix-vector procedure:

$$\mathbf{Y}_{16\times 1} = \mathbf{E}_{16}^{(1)} \mathbf{P}_{16}^{(2)} \mathbf{W}_{16\times 24}^{(1)} \mathbf{P}_{24}^{(4)} \mathbf{W}_{24\times 28}^{(2)} \mathbf{P}_{28}^{(6)} \mathbf{W}_{28\times 30}^{(3)} \mathbf{D}_{30}^{(3)} \widetilde{\mathbf{W}}_{30\times 28}^{(3)} \mathbf{P}_{28}^{(5)} \widetilde{\mathbf{W}}_{28\times 24}^{(2)} \mathbf{P}_{24}^{(3)} \widetilde{\mathbf{W}}_{24\times 16}^{(1)} \mathbf{P}_{16}^{(1)} \mathbf{X}_{16\times 1} \qquad (12)$$

where

$$\mathbf{P}_{28}^{(5)} = \mathbf{I}_{16} \oplus \begin{bmatrix} 1 & & & \\ & 1 & & \\ & & & 1 \\ & & -1 & \end{bmatrix} \oplus \begin{bmatrix} 1 & & & \\ & 1 & & \\ & & & 1 \\ & & -1 & \end{bmatrix} \oplus \begin{bmatrix} 1 & & & \\ & & & 1 \\ & & 1 & \\ & 1 & & \end{bmatrix},$$

$$\mathbf{P}_{28}^{(6)} = \mathbf{I}_{16} \oplus \begin{bmatrix} 1 & & & \\ & 1 & & \\ & & 1 & \\ & & & -1 \end{bmatrix} \oplus \begin{bmatrix} 1 & & & \\ & 1 & & \\ & & 1 & \\ & & & -1 \end{bmatrix} \oplus \begin{bmatrix} 1 & & & \\ & 1 & & \\ & & 1 & \\ & & & -1 \end{bmatrix},$$

$$\mathbf{W}_{28\times 30}^{(3)} = \mathbf{I}_{16} \oplus (\mathbf{I}_2 \otimes (\mathbf{H}_2 \otimes \mathbf{I}_2)) \oplus (\mathbf{T}_{2\times 3} \otimes \mathbf{I}_2), \quad \widetilde{\mathbf{W}}_{30\times 28}^{(3)} = \mathbf{I}_{16} \oplus (\mathbf{I}_2 \otimes (\mathbf{H}_2 \otimes \mathbf{I}_2)) \oplus (\mathbf{T}_{3\times 2} \otimes \mathbf{I}_2),$$

$$\mathbf{D}_{30}^{(3)} = \mathbf{D}_{16}^{(3)} \oplus \mathbf{D}_{14}^{(3)}, \quad \mathbf{D}_{16}^{(3)} = diag\{\tfrac{1}{2}\mathbf{S}_4^{(0)}, \tfrac{1}{2}\mathbf{S}_4^{(1)}, \tfrac{1}{2}\mathbf{S}_4^{(2)}, \tfrac{1}{2}\mathbf{S}_4^{(3)}\}, \quad \mathbf{D}_{14}^{(3)} = diag\{\tfrac{1}{2}\mathbf{S}_2^{(0)}, \tfrac{1}{2}\mathbf{S}_2^{(1)}, \tfrac{1}{2}\mathbf{S}_2^{(2)}, \tfrac{1}{2}\mathbf{S}_2^{(3)}, \mathbf{S}_2^{(4)}, \mathbf{S}_2^{(5)}, \mathbf{F}_2\},$$

$$\mathbf{S}_2^{(0)} = \left[\begin{array}{c|c} -b_4+b_{10}-b_{12}+b_{15} & -b_7-b_9-b_{13}-b_{14} \\ \hline b_7-b_9-b_{13}+b_{14} & -b_4-b_{10}+b_{12}+b_{15} \end{array}\right], \quad \mathbf{S}_2^{(1)} = \left[\begin{array}{c|c} -b_4+b_{10}+b_{12}-b_{15} & -b_7+b_9-b_{13}+b_{14} \\ \hline b_7+b_9-b_{13}-b_{14} & -b_4-b_{10}-b_{12}-b_{15} \end{array}\right],$$

$$\mathbf{S}_2^{(2)} = \left[\begin{array}{c|c} b_4+b_{10}+b_{12}+b_{15} & -b_7-b_9+b_{13}+b_{14} \\ \hline b_7-b_9+b_{13}-b_{14} & b_4-b_{10}-b_{12}+b_{15} \end{array}\right], \quad \mathbf{S}_2^{(3)} = \left[\begin{array}{c|c} b_4+b_{10}-b_{12}-b_{15} & -b_7+b_9+b_{13}-b_{14} \\ \hline b_7+b_9+b_{13}+b_{14} & b_4-b_{10}+b_{12}-b_{15} \end{array}\right],$$

$$\mathbf{S}_2^{(4)} = \left[\begin{array}{c|c} b_9-b_{10} & -b_7-b_{15} \\ \hline -b_7-b_{15} & b_9-b_{10} \end{array}\right], \quad \mathbf{S}_2^{(5)} = \left[\begin{array}{c|c} b_9+b_{10} & -b_7+b_{15} \\ \hline b_7-b_{15} & -(b_9+b_{10}) \end{array}\right].$$

Let us now consider the second order matrices, which were formed as a result of the last decompositions. Matrices $\mathbf{S}_2^{(0)}$, $\mathbf{S}_2^{(1)}$, $\mathbf{S}_2^{(2)}$, $\mathbf{S}_2^{(3)}$, and $\mathbf{F}_2$ can not be factorized, but the matrix $\mathbf{S}_2^{(4)}$ possesses a structure that provides "good" factorization:

$$\mathbf{S}_2^{(4)} = \left[\begin{array}{c|c} b_9-b_{10} & -b_7-b_{15} \\ \hline -b_7-b_{15} & b_9-b_{10} \end{array}\right] = \left[\begin{array}{c|c} a & b \\ \hline b & a \end{array}\right].$$

Then we can write:

$$\mathbf{S}_2^{(4)} = \mathbf{H}_2 \tfrac{1}{2}(s_0 \oplus s_1) \mathbf{H}_2 \qquad (13)$$

Consider now the matrix $\mathbf{S}_2^{(5)}$. If we multiply by (-1) every element of the last row of this matrix, we obtain a new matrix which can be successfully factorized.

$$\widetilde{\mathbf{S}}_2^{(5)} = \left[\begin{array}{c|c} b_9+b_{10} & -b_7+b_{15} \\ \hline -b_7+b_{15} & b_9+b_{10} \end{array}\right],$$

Then we can write:

$$\widetilde{\mathbf{S}}_2^{(5)} = \mathbf{H}_2 \tfrac{1}{2}(s_2 \oplus s_3) \mathbf{H}_2 \qquad (14)$$

Combining all partial factorizations into a single whole and taking into account all manipulations with rows and columns in each matrix at fourth stage of synthesis of algorithm, we obtain a final matrix-vector procedure:

$$\mathbf{Y}_{16\times 1} = \mathbf{E}_{16}^{(1)}\mathbf{P}_{16}^{(2)}\mathbf{W}_{16\times 24}^{(1)}\mathbf{P}_{24}^{(4)}\mathbf{W}_{24\times 28}^{(2)}\mathbf{P}_{28}^{(6)}\mathbf{W}_{28\times 30}^{(3)}\mathbf{P}_{30}\mathbf{W}_{30}^{(4)}\mathbf{D}_{30}^{(3)}\tilde{\mathbf{W}}_{30}^{(4)}\tilde{\mathbf{W}}_{30\times 28}^{(3)}\mathbf{P}_{28}^{(5)}\tilde{\mathbf{W}}_{28\times 24}^{(2)}\mathbf{P}_{24}^{(3)}\tilde{\mathbf{W}}_{24\times 16}^{(1)}\mathbf{P}_{16}^{(1)}\mathbf{X}_{16\times 1} \quad (15)$$

$$\mathbf{W}_{30}^{(4)} = \mathbf{I}_{24} \oplus (\mathbf{I}_2 \otimes \mathbf{H}_2) \oplus \mathbf{I}_2, \quad \tilde{\mathbf{W}}_{30}^{(4)} = \mathbf{I}_{24} \oplus (\mathbf{I}_2 \otimes \mathbf{H}_2) \oplus \mathbf{I}_2, \quad \mathbf{P}_{30} = \mathbf{I}_{27} \oplus (-1) \oplus \mathbf{I}_2,$$

$$\mathbf{D}_{30}^{(3)} = diag(\frac{1}{2}\mathbf{S}_4^{(0)}, \frac{1}{2}\mathbf{S}_4^{(1)}, \frac{1}{2}\mathbf{S}_4^{(2)}, \frac{1}{2}\mathbf{S}_4^{(3)}, \frac{1}{2}\mathbf{S}_2^{(0)}, \frac{1}{2}\mathbf{S}_2^{(1)}, \frac{1}{2}\mathbf{S}_2^{(2)}, \frac{1}{2}\mathbf{S}_2^{(3)}, \frac{1}{2}s_0, \frac{1}{2}s_1, \frac{1}{2}s_2, \frac{1}{2}s_3, \mathbf{F}_2),$$

$$s_0 = -b_7 + b_9 - b_{10} - b_{15}, \quad s_1 = b_7 + b_9 - b_{10} + b_{15}, \quad s_2 = -b_7 + b_9 + b_{10} + b_{15}, \quad s_3 = b_7 + b_9 + b_{10} - b_{15}.$$

Fig. 1 shows a data flow diagram representation of the rationalized algorithm for computation of the Dirac numbers product. In this paper, data flow diagrams are oriented from left to right. Straight lines in the figures denote the operations of data transfer. Points where lines converge denote summation. The dashed lines indicate the sign change operation. We deliberately use the usual lines without arrows on purpose, so as not to clutter the picture. The circles in these figures show the operation of multiplication by a variable (or constant) inscribed inside a circle. In turn, the rectangles indicate the matrix–vector multiplications with the matrix inscribed inside a rectangle.

## 4. Estimation of computational complexity

We calculate how many real multiplications (excluding multiplications by power of two) and real additions are required for realization of the proposed algorithm, and compare it with the number of operations required for a direct evaluation of matrix-vector product in Eq. (2). Let us look to the data flow diagram in Figure 1. It is easy to verify that all the real multiplications which to be performed to computing the product of two Dirac numbers are realized only during multiplying a vector of data by the quasi-diagonal matrix $\mathbf{D}_{30}$. It can be argued that the multiplication of a vector by the matrix $\mathbf{D}_{30}$ requires 88 real multiplications and some trivial multiplications by the power of two. Multiplication by power of two may be implemented using convention arithmetic shift operations, which have simple realization and hence may be neglected during computational complexity estimation.

Now we calculate the number of additions required in the implementation of the algorithm. To count the number of additions required to perform matrix-vector multiplications with matrices $\mathbf{S}_4^{(0)}$, $\mathbf{S}_4^{(1)}$ $\mathbf{S}_4^{(2)}$, and $\mathbf{S}_4^{(3)}$ we introduce the following notation:

$c_{1,1} = b_0 + b_5$, $c_{1,2} = b_{10} + b_{15}$, $c_{1,3} = b_1 + b_2$, $c_{1,4} = b_{13} + b_{14}$, $c_{1,5} = -b_3 + b_4$, $c_{1,6} = -b_{11} + b_{12}$, $c_{1,7} = -b_6 + b_7$, $c_{1,8} = -b_8 + b_9$, $c_{2,1} = b_6 + b_7$, $c_{2,2} = b_8 + b_9$, $c_{2,3} = b_3 + b_4$, $c_{2,4} = b_{11} + b_{12}$, $c_{2,5} = b_1 - b_2$, $c_{2,6} = b_{13} - b_{14}$, $c_{2,7} = b_0 - b_5$, $c_{2,8} = b_{10} - b_{15}$.

Then the matrix $\mathbf{S}_4^{(0)}$ can be represented as follows:

$$\mathbf{S}_4^{(0)} = \begin{bmatrix} c_{1,1} + c_{1,2} & c_{1,3} + c_{1,4} & c_{1,5} + c_{1,6} & c_{1,7} + c_{1,8} \\ -c_{2,1} + c_{2,2} & -c_{2,3} + c_{2,4} & c_{2,5} - c_{2,6} & c_{2,7} - c_{2,8} \\ \hline -c_{2,5} - c_{2,6} & -c_{2,7} - c_{2,8} & c_{1,7} - c_{1,8} & c_{1,5} - c_{1,6} \\ c_{2,3} + c_{2,4} & c_{2,1} + c_{2,2} & c_{1,1} - c_{1,2} & c_{1,3} - c_{1,4} \end{bmatrix}$$

To carry out the multiplication of the matrix $\mathbf{S}_4^{(0)}$ by the corresponding vector we must perform 44 additions, namely:

– 16 additions which are necessary to calculate all the elements $c_{i,j}$, $i, j = 1, 2, ..., 8$,

– 16 additions which are necessary to calculate all the sums $c_{i,j} + c_{i,k}$, $i, j, k = 1, 2, ..., 8$,

– 12 additions arising from the direct matrix-vector multiplication by applying the general rule for matrix-vector multiplication.

Next we estimate the number of real additions that needed for the matrix-vector multiplication with the matrices $\mathbf{S}_4^{(1)}, \mathbf{S}_4^{(2)}, \mathbf{S}_4^{(3)}$:

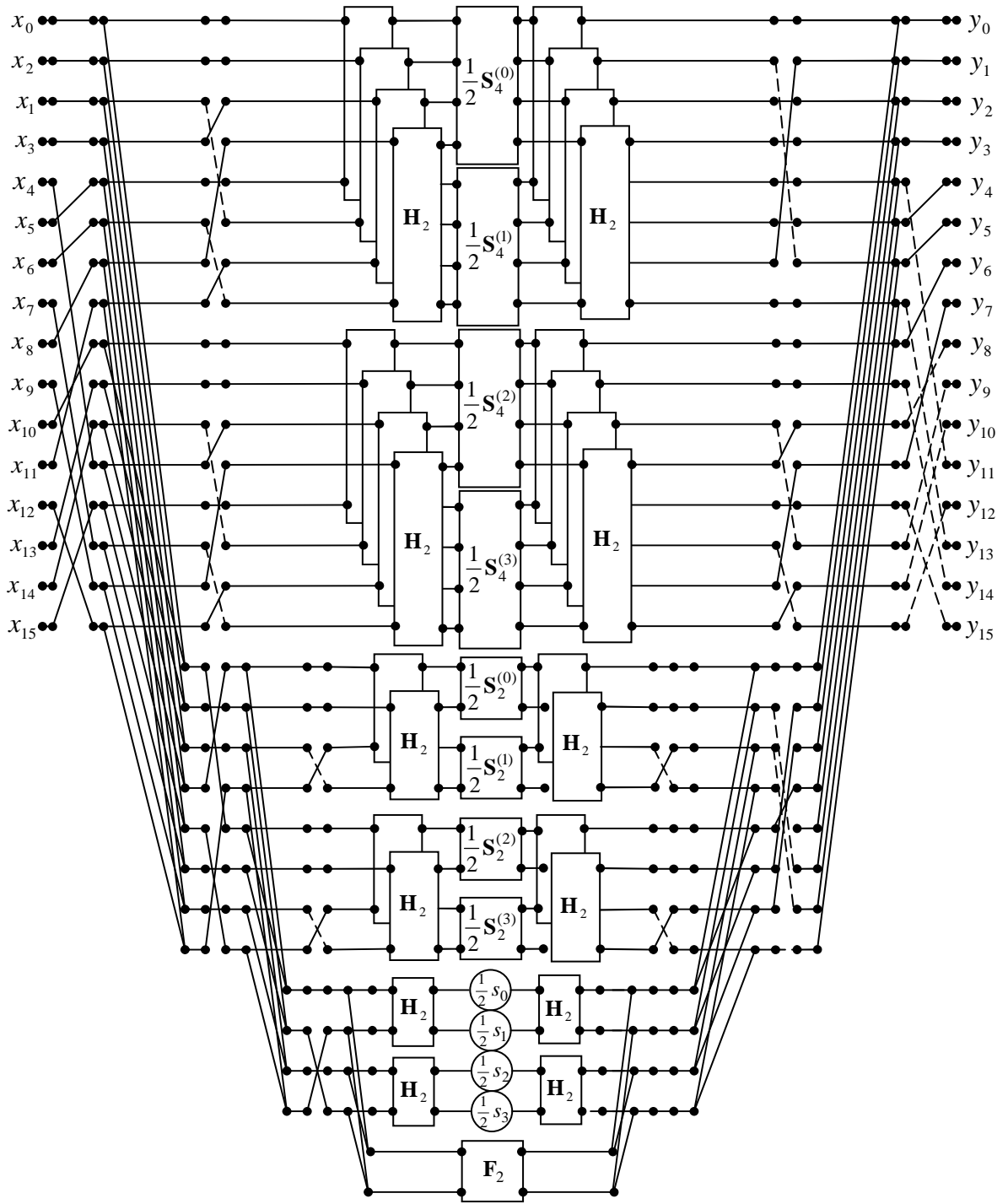

**Fig. 1.** Data flow diagram for rationalized Dirac numbers multiplication algorithm in accordance with the procedure (15).

$$\mathbf{S}_4^{(1)} = \begin{bmatrix} c_{2,7}+c_{2,8} & c_{2,5}+c_{2,6} & c_{1,5}-c_{1,6} & -c_{1,7}-c_{1,8} \\ c_{2,1}+c_{2,2} & c_{2,3}+c_{2,4} & -c_{1,3}+c_{1,4} & c_{1,1}+c_{1,2} \\ c_{1,3}+c_{1,4} & c_{1,1}+c_{1,2} & -c_{1,7}-c_{1,8} & c_{1,5}+c_{1,6} \\ c_{2,3}-c_{2,4} & c_{2,1}-c_{2,2} & c_{2,7}-c_{2,8} & -c_{2,5}+c_{2,6} \end{bmatrix},$$

$$\mathbf{S}_4^{(2)} = \left[\begin{array}{cc|cc} -c_{1,1}+c_{1,2} & -c_{1,3}+c_{1,4} & c_{2,3}+c_{2,4} & c_{2,1}+c_{2,2} \\ -c_{2,5}+c_{2,6} & -c_{2,7}+c_{2,8} & -c_{2,1}+c_{2,2} & -c_{2,3}+c_{2,4} \\ \hline c_{1,5}+c_{1,6} & c_{1,7}+c_{1,8} & -c_{1,1}-c_{1,2} & -c_{1,3}-c_{1,4} \\ -c_{1,7}+c_{1,8} & -c_{1,5}+c_{1,6} & -c_{2,5}-c_{2,6} & -c_{2,7}-c_{2,8} \end{array}\right],$$

$$\mathbf{S}_4^{(3)} = \left[\begin{array}{cc|cc} -c_{2,7}+c_{2,8} & -c_{2,5}+c_{2,6} & c_{2,3}-c_{2,4} & -c_{2,1}+c_{2,2} \\ -c_{1,3}+c_{1,4} & -c_{1,1}+c_{1,2} & -c_{2,1}-c_{2,2} & c_{2,3}+c_{2,4} \\ \hline c_{1,5}-c_{1,6} & c_{1,7}-c_{1,8} & -c_{2,7}-c_{2,8} & c_{2,5}+c_{2,6} \\ c_{1,7}+c_{1,8} & c_{1,5}+c_{1,6} & c_{1,3}+c_{1,4} & -c_{1,1}-c_{1,2} \end{array}\right].$$

In order to implement the multiplication of the matrices $\mathbf{S}_4^{(1)}, \mathbf{S}_4^{(2)}, \mathbf{S}_4^{(3)}$ by the appropriate vectors we need to perform only 28 additions for each of these matrix-vector products because the elements $c_{i,j}$ have already been calculated.

To calculate the number of additions required when performing matrix-vector multiplication with the matrices $\mathbf{S}_2^{(0)}, \mathbf{S}_2^{(1)}, \mathbf{S}_2^{(2)}$, and $\mathbf{S}_2^{(3)}$ we introduce the following notation:

$$p_{1,1} = -b_4 + b_{10}, \; p_{2,1} = b_7 - b_9, \; p_{1,2} = -b_{12} + b_{15}, \; p_{2,2} = -b_{13} + b_{14},$$

$$p_{1,3} = -b_7 - b_9, \; p_{2,3} = -b_4 - b_{10}, \; p_{1,4} = -b_{13} - b_{14}, \; p_{2,4} = b_{12} + b_{15}.$$

Then the matrix $\mathbf{S}_2^{(0)}$ can be represented as follows:

$$\mathbf{S}_2^{(0)} = \left[\begin{array}{c|c} p_{1,1}+p_{1,2} & p_{1,3}+p_{1,4} \\ \hline p_{2,1}+p_{2,2} & p_{2,3}+p_{2,4} \end{array}\right]$$

In order to carry out the multiplication of the matrix $\mathbf{S}_2^{(0)}$ by an appropriate vector we need to perform only 14 additions, namely:

– 8 additions which are necessary to calculate all the elements $p_{i,j}$, $i,j = 1,2,3,4$,

– 4 additions which are necessary to calculate all the sums $p_{i,j} + p_{i,j+1}$, $i,j = 1,2,3,4$.

– 2 additions arising from the direct matrix-vector multiplication by applying the general rule for matrix-vector multiplication.

In order to carry out the multiplication of the matrices $\mathbf{S}_2^{(1)}, \mathbf{S}_2^{(2)}, \mathbf{S}_2^{(3)}$ by appropriate vectors we need to perform only 6 additions for each of these matrix-vector products because the elements $p_{i,j}$ have already been calculated:

$$\mathbf{S}_2^{(1)} = \left[\begin{array}{c|c} p_{1,1}-p_{1,2} & -p_{2,1}+p_{2,2} \\ \hline -p_{1,3}+p_{1,4} & p_{2,3}-p_{2,4} \end{array}\right], \; \mathbf{S}_2^{(2)} = \left[\begin{array}{c|c} -p_{2,3}+p_{2,4} & p_{1,3}-p_{1,4} \\ \hline p_{2,1}-p_{2,2} & -p_{1,1}+p_{1,2} \end{array}\right], \; \mathbf{S}_2^{(3)} = \left[\begin{array}{c|c} -p_{2,3}-p_{2,4} & -p_{2,1}-p_{2,2} \\ \hline -p_{1,3}-p_{1,4} & -p_{1,1}-p_{1,2} \end{array}\right].$$

To calculate elements $s_0, s_1, s_2, s_3$ we need to perform only 4 additions because

$$s_0 = -b_7 + b_9 - b_{10} - b_{15} = -(b_7 - b_9) - (b_{10} + b_{15}) = -p_{21} - c_{12},$$

$$s_1 = b_7 + b_9 - b_{10} + b_{15} = (b_7 + b_9) - (b_{10} - b_{15}) = -p_{13} - c_{28},$$

$$s_2 = -b_7 + b_9 + b_{10} + b_{15} = -(b_7 - b_9) + (b_{10} + b_{15}) = -p_{21} + c_{12},$$

$$s_3 = (b_7 + b_9) + (b_{10} - b_{15}) = -p_{13} + c_{28}.$$

In order to carry out the multiplication of the matrix $\mathbf{F}_2$ by an appropriate subvector we need to perform 2 additions.

Thus for multiplying the data vector by the quasi-diagonal matrix $\mathbf{D}_{30}$, we need to perform 88 multiplications and 166 additions. The Fig. 1 shows that the implementation of the remaining part of the algorithm requires only 90 additions. Thus using the proposed algorithm the number of real multiplications to calculate the Dirac number product is reduced threefold compared to schoolbook method of calculation. The number of real additions required using our algorithm is 256. Therefore, the total number of arithmetic operations for proposed algorithm is approximately 30% less than that of the direct evaluation.

## 5. Conclusion

In this paper, we have presented an original algorithm that allows us to compute the product of two Dirac numbers with reduced multiplicative complexity. The proposed algorithm saves 40 real multiplications compared to the algorithm [12] and 168 real multiplications compared to the schoolbook algorithm. Unfortunately, the number of real additions in the proposed algorithm is somewhat greater than in the algorithm [12], but the total number of arithmetical operations is still less than in the schoolbook algorithm. For applications where the "cost" of a real multiplication is greater than that of a real addition, the new algorithm is always more computationally efficient than our previously published algorithm, and it is generally more efficient than direct method.